\documentclass[sigconf]{acmart}

\usepackage{pifont} 

\usepackage{hyperref}
\usepackage{float}

\usepackage{multirow}
\usepackage{caption}
\usepackage{longtable}
\usepackage{placeins} 
\usepackage{dblfloatfix} 

\usepackage{colortbl}
\usepackage{makecell}
\usepackage{graphicx} 

\usepackage{booktabs}      
\usepackage{array}         

\copyrightyear{2026}
\acmYear{2026}
\setcopyright{cc}
\setcctype{by}
\acmConference[CAIN '26]{2026 IEEE/ACM 5th International Conference on AI Engineering - Software Engineering for AI}{April 12--13, 2026}{Rio de Janeiro, Brazil}
\acmBooktitle{2026 IEEE/ACM 5th International Conference on AI Engineering - Software Engineering for AI (CAIN '26), April 12--13, 2026, Rio de Janeiro, Brazil}
\acmDOI{10.1145/3793653.3793785}
\acmISBN{979-8-4007-2475-6/2026/04}

\begin{document}

%
\title{A Systematic Review of MLOps Tools: Tool Adoption, Lifecycle Coverage, and Critical Insights}

%


\author{Zakkarija Micallef}
\affiliation{
  \institution{Vrije Universiteit Amsterdam}
  \country{The Netherlands}
}
\email{zak.micallef@proton.me}
\author{Keerthiga Rajenthiram}
\affiliation{
  \institution{Vrije Universiteit Amsterdam}
  \country{The Netherlands}
}
\email{k.rajenthiram@vu.nl}
\author{Ilias Gerostathopoulos}

\affiliation{
  \institution{Vrije Universiteit Amsterdam}
  \country{The Netherlands}
}
\email{i.g.gerostathopoulos@vu.nl}


%

%

\begin{abstract}
Machine Learning Operations (MLOps) has become increasingly critical as more organisations move ML models into production. However, the growing landscape of MLOps solutions has introduced complexity for practitioners trying to select appropriate tools. To investigate how and why these tools are adopted in practice, this paper conducts a systematic review of the academic literature focused on MLOps tools. We map tools to MLOps lifecycle components to reveal their function, scope, and the challenges they are designed to address. We identify usage trends and synthesise reported benefits and limitations. The most commonly used components, according to the findings, are orchestration frameworks, data versioning, experiment tracking, and managed cloud platforms. No single tool covers the entire lifecycle, so researchers often combine multiple tools to build complete pipelines. This highlights the importance of interoperability across MLOps tools in real-world MLOps pipelines.
\end{abstract}

%
%



%


%

%
\maketitle

\section{Introduction}
In recent years, AI has experienced a dramatic surge in popularity. As more companies deploy AI solutions, they quickly discover that specialised infrastructure is essential to support these new capabilities. However, many AI engineers and data scientists lack the software-deployment expertise of operations teams, especially when it comes to Machine Learning Operations (MLOps) tools \cite{intro_1_popular_ai_Davenport2021Deployment}. 
 

MLOps refers to the entire lifecycle of the machine learning (ML) process, bridging the gap between data, development, and operations \cite{kreuzberger_machine_2022}. MLOps extends beyond applying DevOps principles \cite{2_devops} to ML. It involves continuous integration and continuous deployment automation for ML pipelines, orchestration of ML workflows, versioning of data, models, and code to ensure reproducibility, continuous training to keep models up to date, metadata tracking for experiment auditability, and continuous evaluation and performance monitoring.

Despite the growing interest in MLOps, existing reviews often remain high-level, focusing primarily on listing tools, comparing surface-level features, or distinguishing between open-source and proprietary solutions~\cite{stone_navigating_2025,symeonidis_mlops_2022,testi_mlops_2022,berberi_machine_2025}. However, there is limited synthesis of how these tools are actually used in practice \cite{recupito_multivocal_2022}. Most studies fail to capture the practical experiences of teams that deploy real-world ML systems. This gap in the literature limits the ability of researchers and practitioners to make informed decisions based on implementation outcomes rather than tool specifications alone. 

In this paper, we perform a Systematic Literature Review (SLR) to explore the range of MLOps tools referenced in the academic literature and examine their real-world applications. While other papers integrate general-purpose DevOps tools such as Docker, Kubernetes, Git, and Jenkins with ML-specific tools \cite{faezeh_sms}, our study focuses exclusively on MLOps native tools that are developed specifically for ML workloads, excluding general-purpose tools. We explore how practitioners use these tools in their pipelines, why they select specific MLOps solutions, and the advantages and drawbacks they face in real-world scenarios. Through this analysis, we aim to provide a clear view of the current MLOps landscape.

\section{Related Work}
\label{sec:related_work}


MLOps remains a vague umbrella term, with its scope and implications still ambiguous for both researchers and practitioners. To bring clarity, Kreuzberger et al. \cite{kreuzberger_machine_2022} conducted mixed‐method research, where they combined a literature review, a tool review, and expert interviews to create a comprehensive description of MLOps that we adopt as our definition of MLOps in Section 1. Their work has become a \textit{de facto} standard, widely cited by other papers \cite{bodor_mlops_2023} \cite{wazir_mlops_2023} \cite{zarour_mlops_2025}. While Kreuzberger \textit{et al.} \cite{kreuzberger_machine_2022} identified principles, technical components, and roles of MLOps, we focus solely on the technical elements that recur across the literature
and
form the backbone of MLOps solutions \cite{recupito_multivocal_2022} \cite{wazir_mlops_2023} \cite{zarour_mlops_2025} \cite{bodor_mlops_2023} \cite{faezeh_sms}:
(i) Data engineering
(ii) Version control 
(iii) Hyperparameter tuning and experiment tracking
(iv) CI/CD pipelines
(v) Workflow orchestration 
(vi) Model deployment/serving
(vii) Automated testing and validation
(viii) Continuous performance monitoring.
%


Varon Maya \cite{maya_state} argues that bringing DevOps principles into ML pushes organisations toward pipeline-like architectures: NVIDIA \cite{nvidia}, Facebook \cite{facebook}, Spotify \cite{spotify} and Google \cite{google} each describe a flow that runs from data collection and feature engineering through training, validation and model serving, occasionally adding feedback loops that trigger continuous retraining. This structure clarifies ownership and lets every stage use specialised tooling. 
Our systematic literature review (SLR) adopts the same pipeline perspective, evaluating MLOps tools stage by stage. 


In Recupito et al.'s review \cite{recupito_multivocal_2022}, thirteen prominent MLOps platforms were mapped to the stage of the MLOps pipeline they support: data management, model training, CI/CD, monitoring, and so on. Their comparison shows that no single product spans the entire lifecycle, so practitioners routinely assemble multi-tool pipelines. Multiple studies \cite{recupito_multivocal_2022} \cite{ruf_demystifying_2021} \cite{wazir_mlops_2023} highlight the importance of evaluating not just each tool's individual strengths but also how well they interoperate with other tools and the dependencies they introduce. 

Our SLR compares tools by the benefits and limitations that paper authors explicitly report from their practical experience, in contrast to the majority of previous reviews that compare tools based on claimed features listed on the tool developer's websites or their official blogs. This method provides a more accurate understanding of the practical experience and shortcomings of MLOps tools.
\section{Study Design}\label{sec:design}
\label{sec:design}


We provide here an overview of our study design; for full replicability we refer the interested reader to the study's online appendix~\cite{Micallef2025_MLOps_Appendix}.

\paragraph{Research Questions}
\label{sec:questions}

Our study aims to answer the following three research questions (RQs).

\begin{itemize}
    \item[RQ1] \textit{Which tools employed in MLOps workflows are most frequently reported in academic literature?}
    The goal of this research question is to determine which tools have been the subject of academic investigation most frequently. 
    Tools with significant popularity benefit from strong community support and rich integration possibilities, making them preferred candidates for further exploration.
    \item[RQ2] \textit{Which components of the MLOps lifecycle do these tools cover in the use cases reported in the academic literature?}
    Analyzing which MLOps components each tool covers reveals its function, scope and challenges it is designed for. This understanding is crucial since, MLOps engineers tend to combine multiple tools to get their final desired pipeline.
    \item[RQ3] \textit{What are the reported benefits and limitations of the MLOps tools?}
    MLOps tools offer a variety of features designed to meet diverse needs. Following the analysis of existing literature and documented user experiences, we examine common trends in their advantages and limitations.  
\end{itemize}



\paragraph{Initial Search}

We conduct our literature search on Google Scholar in February 2025. Our search string combines relevant terms from known MLOps literature with synonyms:

\textit{\noindent "MLOps" OR "machine learning operations") AND ("tool" OR "application" OR "framework"  OR "platform" OR "pipelines") AND ("comparison" OR "evaluation" OR "benchmark" OR "analysis" OR "empirical"}


The query returns 9,100 records. To keep the effort manageable, we screen the first 10 pages (96 papers) for inclusion. After applying our selection criteria, 27 articles were included. 





\paragraph{Selection Criteria} \label{criteria}

We apply the following inclusion (IC) and exclusion (EC) criteria, refined over a 10-paper pilot study.

\begin{itemize}
    \item[IC1] Analyzes/compares MLOps tools or describes the development/implementation of MLOps tools, frameworks, or pipelines.
    \item[IC2] Published within the last five years (2020 or later).

    \item[EC1] Lacks detailed tool analysis; only discusses applications or projects that use an MLOps tool.
    \item[EC2] Discusses MLOps only as a concept/architecture or highlights the benefits of the MLOps culture.
    \item[EC3] It is a secondary study, survey, or book.
    \item[EC4] It is not written in English.
    \item[EC5] The full paper is not available for our institution.
\end{itemize}



IC2, EC4, and EC5 are standard criteria for SLRs. IC1, EC1-EC3 ensure that we exclude papers lacking original primary research or detailed analysis/development of MLOps tools and thus focus only on studies that directly contribute to our research questions.

\paragraph{Snowballing}

We apply both backward and forward snowballing to expand the results obtained through our initial search, following the guidelines of Wohlin et al. \cite{wohlin_guidelines_2014}. 
The process yields a total of 14 additional studies (2 from backward and 12 from forward snowballing).
We limit the snowballing process to a single round since further iterations yields minimal additional relevant papers.




\paragraph{Data Extraction} \label{extract}


Table \ref{tab:extraction-fields} lists the data extraction fields we use and their mapping to the RQs.
For RQ1, we record tools that are actively implemented or used, excluding tools that are simnly mentioned. This enables us to evaluate the researchers' hands-on experience with the tools. For RQ2, we note the tools' specific use cases and MLOps stages they cover as mentioned in the papers. Finally, for RQ3, we collect the reported benefits and limitations of each tool by analysing the justifications for their selection and the challenges encountered during their application.

\begin{table}[ht]
\vspace{-0.2cm}
\small
  \centering
  \caption{Data extraction fields}
  \label{tab:extraction-fields}
  \begin{tabular}{lll}
    \toprule
    \textbf{Field} & \textbf{RQ}             & \textbf{Description}                                    \\
    \midrule
    Name          & RQ1              & Name of the MLOps tool                                  \\
    Uses          &    RQ2           & Practitioner use cases                                  \\
    Pipeline stage   & RQ2     & Stage(s) of the MLOps lifecycle addressed               \\
    Benefits          & RQ3           & Reported benefits of the tool                           \\
    Limitations        & RQ3          & Reported limitations or challenges                      \\
    \bottomrule
  \end{tabular}
\end{table}

\paragraph{Data Synthesis}

For RQ1, we construct a frequency graph that captures the popularity of each tool. 
For RQ2, we categorize the tools according to their respective use cases and stages, according to the MLOps components defined by Najafabadi \textit{et al.}~\cite{faezeh_sms}.
For RQ3, we apply open coding to synthesize benefits or limitations out of the extracted paper fragments.
The categorization and coding is performed by the first author and reviewed by the other authors.

\paragraph{Threats to Validity}

Our study is subject to the following validity threats~\cite{validity}. 
With respect to \textbf{external validity}, i.e. the generalizability of our findings, our decision to review only the first ten pages of Google Scholar results means that we may have missed some relevant studies. 
While Scholar's proprietary relevance ranking does act as a partial mitigation by prioritizing influential papers first, a broader screening would reduce this threat further. 
A threat to \textbf{internal validity} concerns tool–component heatmap (Figure \ref{fig:heatmap}). We can map only the capabilities that the authors explicitly describe, so the heatmap may miss components that a tool supports, but are not used in the included studies. 
A threat to \textbf{construct validity} is inherent in the source literature: many academic papers emphasise implementing MLOps pipelines or addressing broader issues rather than critically evaluating the tools themselves. This tendency often leads to detailed reporting on the benefits while underreporting limitations, which may result in conclusions that do not accurately capture the tool's effectiveness.
Finally, with respect to \textbf{conclusion validity}, our conclusions, based on tools occurrences, component mappings, and feature evaluations, are meant to be rational. However, since the extraction and synthesis process was not reviewed by a third party, our results are still prone to self-bias.

\section{Results} \label{sec:results}

\subsection{RQ1: Which tools employed in MLOps workflows are most frequently reported in academic literature?} \label{results-rq1}

\begin{figure}[t]
    \centering
    \includegraphics[width=0.9\linewidth]{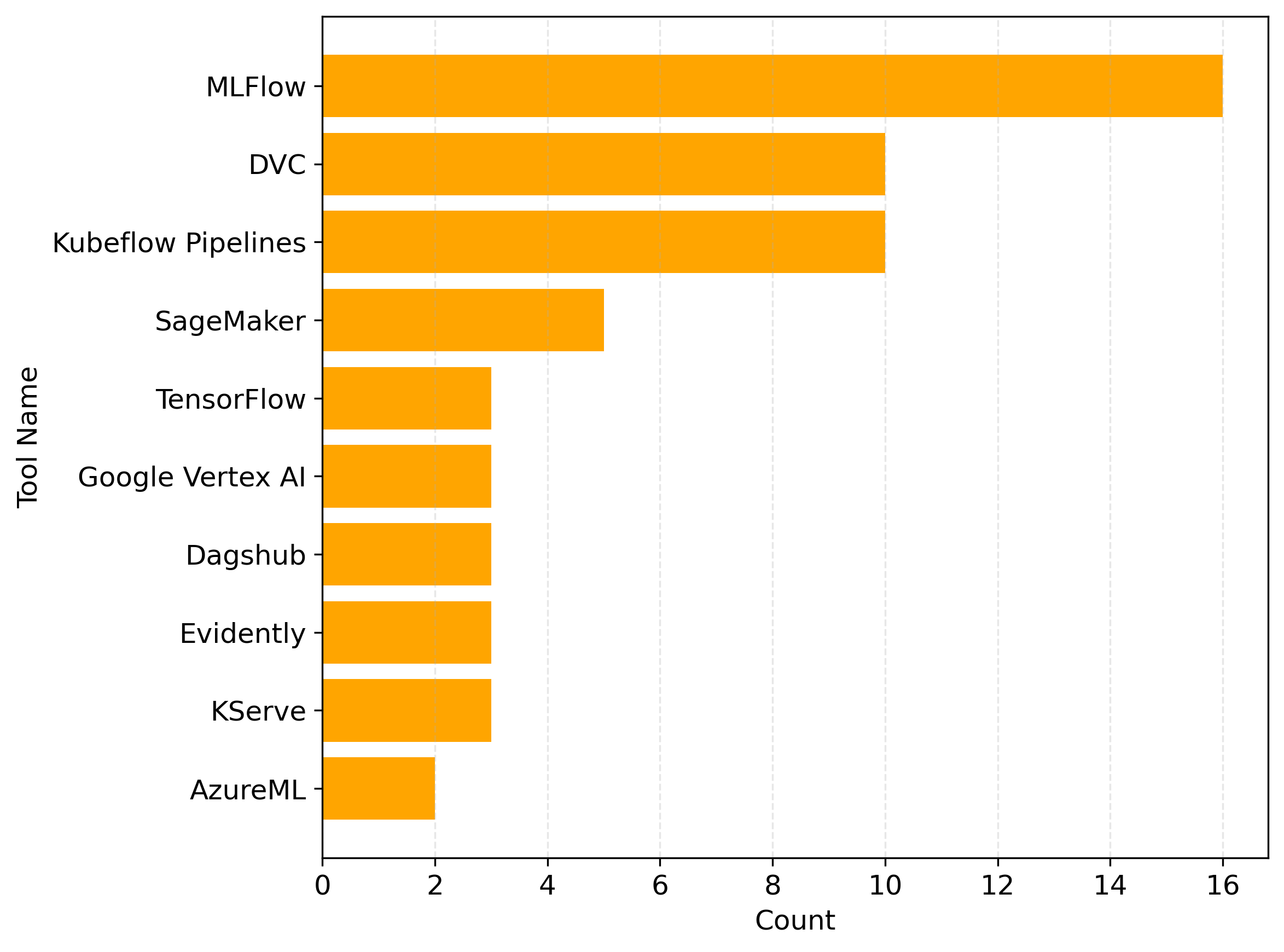}
    \caption{Top 10 MLOps tools ranked by number of mentions.
    }
    \label{fig:count}
\end{figure}

A total of 31 MLOps tools were identified across our 41 included primary studies. 
Figure \ref{fig:count} shows the prevalent tools.

Across the primary studies, four tools stand out: \textbf{MLflow}, \textbf{DVC}, \textbf{Kubeflow Pipelines}, and \textbf{AWS SageMaker} as shown in Figure \ref{fig:count}. The first three are fully open-source, while AWS SageMaker is a proprietary cloud service. This divide reflects a familiar compromise in MLOps practice: practitioners prefer community-maintained tools for tasks like experiment tracking and data versioning but often turn to commercial platforms when they need infrastructure that scales quickly and comes with operational support.

Both MLflow and Kubeflow are classified as end-to-end platforms, providing experiment tracking, model packaging, and deployment in one bundle. However, their design philosophies diverge significantly. MLflow focuses on accessibility: it is easy to use, has a straightforward web UI, thorough and substantial documentation, wide storage-backend support, and a lively community \cite{1}\cite{3}\cite{5}\cite{6}\cite{8}\cite{16}\cite{22}\cite{24}\cite{38}\cite{39}. Kubeflow takes a different approach. It is rooted in Kubernetes and focuses more on customisability with fine-grained control, modular components, and automatic scaling \cite{1}\cite{13}\cite{29}\cite{31}\cite{35}\cite{36}\cite{41}. However, it demands a tougher setup and steeper operational know-how \cite{27}\cite{41}. The fact that both tools top the frequency chart suggests that while some users prefer easy-to-use tools and others value customisable options, there is room for both approaches to succeed.

Another reason for MLflow's popularity is that, even though it is open-source, it remains a mature, well-established platform trusted by leading companies and backed by both a strong community and Databricks \cite{16}\cite{38}. Its popularity creates a positive feedback loop: more users attract more contributors, the codebase improves, and the project becomes even more appealing, which leads the community to further grow. The same holds true for Kubeflow \cite{36}. In contrast, tools like Polyaxon face "uncertain vitality" due to smaller contributor bases \cite{41}. Active maintainer communities weigh heavily in long-term tool selection \cite{16}\cite{36}\cite{38}. MLflow does not just entice developers who want a self-hosted solution, since developers who prefer to skip infrastructure work can opt for a managed MLflow provided by Databricks, which sidesteps the need to run and maintain a separate tool \cite{5}\cite{6}.

DVC is a notable outlier among the most cited tools, as it tackles only a single stage of the MLOps lifecycle, data storage and versioning, and yet it appears almost as often as the full end-to-end platforms. Synchronising large data artefacts with code has been a historic pain point in ML. Authors repeatedly select DVC for its easy-to-use Git-like interface \cite{5}, pipeline caching and serverless architecture. A recurring pairing was the adoption of Git for code with DVC for large data because of Git's size limits. Git LFS was meant to alleviate this struggle, but unlike Git LFS, DVC needs no extra server, a decisive difference 
\cite{5}. 
It bears noting that DVC's full feature set extends to experiment tracking but per our methodology, we map only features explicitly discussed 
in the literature.

The widespread use of AWS SageMaker highlights the ongoing significance of managed services (Figure \ref{fig:count}). Managed MLOps platforms such as SageMaker remain popular because they include a wide array of pre-integrated pipeline components, such as a feature store, model registry, and CI/CD templates \cite{23}\cite{33}. Companies accept the subscription fee and the inherent risks of vendor lock-in because the alternatives present significant challenges and costs \cite{21}\cite{33}. A managed service hides the heavy work of provisioning and securing compute, storage, and networking, and it scales on demand \cite{7}\cite{28}\cite{33}. The same attraction applies to the managed stacks from Azure ML \cite{33} and Google Cloud Vertex AI \cite{7}. Organisations that are unwilling to rely on a third-party cloud provider often choose to build an in-house pipeline from open-source projects such as MLflow for experiment tracking, Kubeflow for orchestration, and DVC for data versioning. However, this option demands far more integration effort and long-term maintenance and may be better suited to larger teams and more mature projects \cite{5}\cite{6}\cite{16}\cite{24}\cite{27}\cite{41}.

\subsection{RQ2: Which components of the MLOps lifecycle do these tools cover in the use cases reported in the academic literature?} \label{results-rq2}

MLOps pipelines encompass a range of components, including data ingestion, pre-processing, model development, training, validation, deployment, and ongoing monitoring.
Since MLOps tools can address several components, authors may discuss only a subset. As a result, the figure may omit certain components a tool could cover, making our assessment non-exhaustive.

\begin{figure*}[!t]
    \centering
    \includegraphics[width=0.83\textwidth]{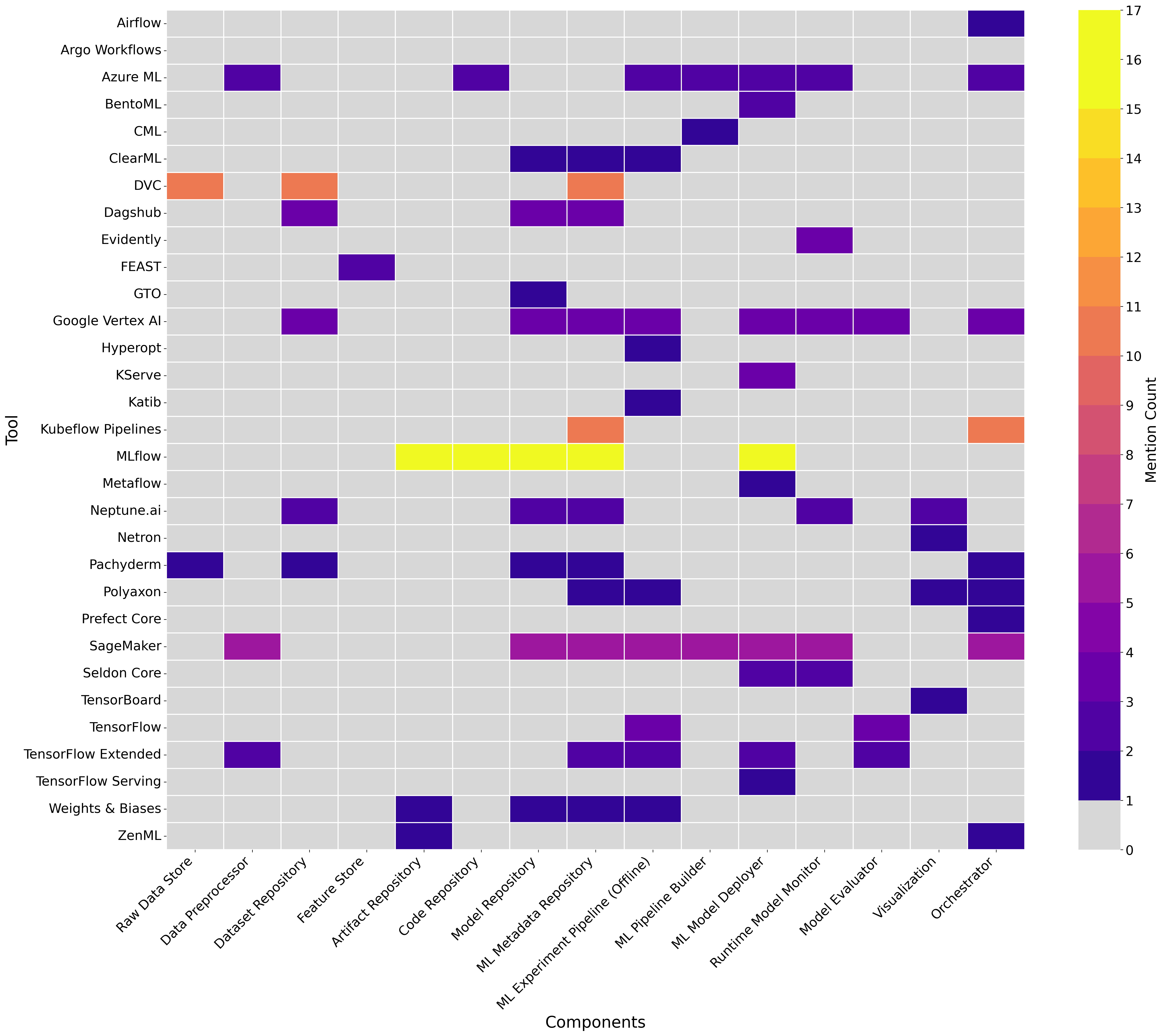}
    \caption{Heatmap of MLOps tools mapped to their corresponding pipeline components adapted from Najafabadi et al.\cite{faezeh_sms}. See Table 1 in online appendix~\cite{Micallef2025_MLOps_Appendix} for a description of each component. 
    }
    \label{fig:heatmap}
\end{figure*}



Mapping each tool to categories from Najafabadi \emph{et al.}'s taxonomy \cite{faezeh_sms} (Figure \ref{fig:heatmap}) reveals that no single solution addresses the entire ML lifecycle. However, this is due more to what the papers mention than to what the products definitively offer. Kubeflow excels at orchestration \cite{36}, MLflow at experiment tracking \cite{1}\cite{3}\cite{5}, while DVC and Feast handle data and feature management, respectively \cite{5}\cite{20}.

AWS SageMaker bundles a model registry, feature store, and deployment tools, yet teams still turn to third-party services for granular security and local runs \cite{21}\cite{33}. The component that offers the least coverage is in feature stores and runtime monitoring; outside of Feast and Evidently, non-managed options are almost nonexistent. Consequently, multi-tool pipelines are the norm, underpinning the importance of effective tool integration.
While classifying tools, we found one component that did not fit cleanly into Najafabadi \textit{et al.'s} \cite{faezeh_sms} taxonomy: visualisation dashboards such as TensorBoard and Weights \& Biases so we introduced a dedicated visualisation component. Across all reviewed papers, the ML Metadata Repository (Experiment-tracking) stage is addressed the most. This dominance is caused by the ubiquity of MLflow, making it a default choice for an experiment tracking tool that is open source. Managed solutions such as SageMaker and Azure ML \cite{23}\cite{33} further contribute to its widespread adoption. Reliable metadata is essential for reproducibility, auditability, and model evaluation; consequently, practitioners consistently prioritise this component. Orchestration tools sit a close second. They bind data prep, training, and deployment into a runnable pipeline, so virtually every study that examines end-to-end workflows also highlights orchestrators such as Kubeflow, Argo, or similar schedulers.

\subsection{RQ3: What are the reported benefits and limitations of the MLOps tools?} \label{results-rq3}

We provide here a thematic analysis of the findings of RQ3. For the full benefits and limitations per tool we refer the interested reader to Tables 2-9 in the online appendix~\cite{Micallef2025_MLOps_Appendix}.

\paragraph{Security and Managed Platform Trade-offs} \label{results-rq3-security}
MLflow lacks automatic data versioning \cite{5}\cite{24}, which explains why many studies pair it with DVC \cite{5}\cite{14}\cite{15}. It is also missing role-based access control \cite{16}\cite{24}. In contrast, AWS SageMaker covers that gap through IAM integration \cite{23}\cite{33}. A well‐established trade‐off emerges: open-source tools need extra engineering to meet enterprise-grade security, while cloud platforms shift that work to the provider at the cost of vendor lock-in. Even then, the cloud does not guarantee full coverage. For example, Google's Vertex AI still trails Azure ML and SageMaker on security, user control, and governance despite tying you to Google's stack.

\paragraph{Post-deployment Monitoring}
Evidently is the only tool that comes with a drift detection feature out of the box \cite{4}\cite{20}\cite{31}, and it works only for tabular data \cite{31}. The limited findings indicate that post-deployment observability is still an emerging component.

\paragraph{Prerequisite Knowledge}
A common limitation among tools is the required prerequisite knowledge. Pachyderm requires Helm and cloud-storage expertise \cite{41}, while Kubeflow and Argo Workflows demand solid Kubernetes and containerisation experience \cite{27}\cite{41}. Other services, including Feast's SDK \cite{21}, Weights \& Biases' client code \cite{2}\cite{16}, Neptune's Python API \cite{2} and SageMaker's SDK \cite{21}, call for advanced programming skills. These findings show that tool comparison should not focus solely on a tool's features and limitations, but it must also consider the skills a team possesses.

\paragraph{Integration and Flexibility}
Easy integration and language agnosticism are among the most frequently praised benefits. Orchestrators such as Kubeflow Pipelines are applauded for both their cloud-agnostic Kubernetes foundation and their seamless hooks into TensorFlow Extended \cite{1}\cite{29}\cite{31}\cite{36}\cite{41}, while managed stacks win favour largely because of the way they slot into their parent ecosystems such as AWS SageMaker's seamless integration with the rest of the AWS \cite{23}\cite{33}. At the experiment-tracking layer, MLflow extends this integrative spirit through container-friendly, self-hosted deployments and pluggable back-end stores that work just as well with S3, Azure Blob, on-prem NFS, or SQL-compatible databases \cite{6}\cite{16}\cite{24}. Flexibility in storage backends is also demonstrated by DVC's remote options \cite{5} and MLflow's broad object-store support \cite{24}. TFX is widely appreciated for its portability, with pipelines that can run seamlessly across multiple orchestrators rather than being locked to a single workflow engine \cite{7}. Finally, inference services such as BentoML embrace framework diversity by supporting TensorFlow, PyTorch, Keras, XGBoost, and more out of the box \cite{6}\cite{9}. Taken together, these examples show that the community consistently rewards tools that can drop into existing tech stacks without forcing a wholesale rewrite to a particular technology. In contrast, the lack of flexibility is listed as a drawback by reviewers, specifically in TensorBoard and TFX which tie users to the TensorFlow stack 
\cite{21}\cite{36}.

\paragraph{User Interfaces and Visualization} \label{results-rq3-ui}
Finally, visual dashboards and UIs are widely appreciated. Kubeflow's central UI \cite{13}\cite{41}, Vertex AI's pipeline view \cite{7}, MLflow's experiment board \cite{3}\cite{8}\cite{24} and Pachyderm's web console \cite{41} are all reported as benefits, as they improve ease of use.

\section{Discussion} \label{sec:discussion}

\paragraph{Implications for Practitioners}


Our data shows that no single tool covers the entire MLOps lifecycle, making tool integration and compatibility a must. Practitioners should allocate additional time and resources for integration work and to confirm the interoperability of chosen tools. Practitioners must also consider the trade-off between open-source flexibility and managed-service convenience. Building a pipeline from open-source components offers greater control and customization \cite{6}\cite{16}\cite{24} but requires significant integration effort and ongoing maintenance \cite{5}\cite{6}\cite{16}\cite{24}\cite{27}\cite{41}. Conversely, managed services reduce both integration and operational overhead \cite{11}\cite{23}\cite{33} but introduce vendor lock-in risks \cite{33} and subscription costs \cite{21}\cite{33}. Our findings unexpectedly highlight the need to consider team skillsets. The prerequisite knowledge requirements identified in our review range from Kubernetes expertise for Kubeflow \cite{27}\cite{41}, to containerization skills for Argo Workflows, and cloud storage knowledge for Pachyderm \cite{41}, which some practitioners reported as a major blocker. 
Community vitality also proved crucial: tools like MLflow and Kubeflow benefit from active open-source communities providing comprehensive documentation, integrations, and long-term support \cite{16}\cite{36}\cite{38}, whereas tools with limited contributor bases, such as Polyaxon, face ``uncertain vitality'' \cite{41} that may risk long-term sustainability.

\paragraph{Implications for Researchers}
Our systematic review indirectly tested Najafabadi et al.~\cite{faezeh_sms}'s taxonomy, finding it largely effective for categorizing MLOps tools. A notable gap emerged in visualization, with tools like TensorBoard and Weights \& Biases not fitting cleanly into any existing component. We also introduced the categories ``End-to-End'' and ``Managed End-to-End'' to enable higher-level analysis. These changes highlight that such taxonomy requires continuous refinement as the MLOps ecosystem evolves.

A critical research gap emerged from our analysis. Although our data confirm that practitioners routinely combine multiple tools to create complete MLOps pipelines, the literature lacks measures and studies that focus on integration effort and complexity.
\FloatBarrier  

\section{Conclusion}

This systematic review of 41 academic studies reveals that MLflow, DVC, Kubeflow Pipelines, and AWS SageMaker dominate current MLOps practice, each addressing a critical pain point. Their popularity is sustained either by lively open-source communities or by the deep resources of a major cloud provider. However, no single tool covers the entire ML lifecycle, forcing practitioners to assemble multi-tool pipelines. Future work should survey grey literature to validate whether the tool usage patterns observed in academic literature align with industry practice. Additionally, empirical studies quantifying the integration effort for common tool combinations (e.g., MLflow with DVC, Kubeflow with Feast) would provide practitioners with actionable metrics for pipeline planning.

\begin{acks}
This research is supported by ExtremeXP, a project co-funded by the European Union Horizon Programme under Grant Agreement No. 101093164.
\end{acks}

\bibliographystyle{ACM-Reference-Format}
\bibliography{bibliography}

%

\end{document}